\documentclass[twocolumn,prb,aps,showpacs,groupedaddress]{revtex4}
\usepackage{array}
\usepackage{graphicx}

\def\be{\begin{equation}}
\def\ee{\end{equation}}
\def\bea{\begin{eqnarray}}
\def\eea{\end{eqnarray}}
\def\ba{\begin{array}}
\def\ea{\end{array}}
\def\bdm{\begin{displaymath}}
\def\edm{\end{displaymath}}

\begin{document}

\title{Kinetic equation and magneto-conductance for Weyl metal in the clean limit}

\author{S.-K. Yip}

\affiliation{Institute of Physics and Institute of Atomic and Molecular Sciences,
 Academia Sinica, Taipei, Taiwan}

\date{\today }

\begin{abstract}

We discuss the semi-classical kinetic equation in the clean limit,
 with the presence of Berry curvatures and
magnetic field $B$, with the aim of applying to Weyl semi-metals.  Special
attention is given to the conservation laws for the collision integrals.
It is found that the magneto-resistance second order in $B$
is in general negative, with or without Weyl points, though
in the later case it is in general much smaller.

\end{abstract}

\pacs{
72.10.Bg, 	
73.43.Qt, 	
75.47.-m 	
}

\maketitle

\section{Introduction}

Recently, there are a lot of attentions on magneto-transport in
Weyl semi-metals. \cite{Kim13,Zhang15,Huang15,Xiong15,Li15}
In this paper, I consider the clean limit kinetic equation
for an electronic system in the presence of Berry
curvatures at low magnetic fields,
in order to better understand transport in these
systems.
There are previous related efforts.
There is a paper by Son and Spivak \cite{SS13}, extending earlier work of
\cite{NM83} and concentrating  on chiral anomaly.
They therefore consider the case where
there are two (or more) Fermi pockets (valleys), with each pocket enclosing a separate
Weyl point.   Furthermore, they assumed that there is a separation of
collision time scales, with the intra-pocket scattering time negligible compared
with inter-pocket ones.  They then deduce the magneto-conductivity in this
limit with the result depending entirely on inter-pocket scattering times.
 In Kim et al \cite{Kim13} (see
in particular the Supplementary Material), transport equations were derived
first using a Green's function approach, then a kinetic equation.
The first approach is complicated and the results not easy to use or interpret.
In their derivation for the kinetic equation, the Fermi exclusion principle
was ignored, making the validity of results difficult to judge.
Subsequently, some of the authors of \cite{Kim13} provided another
derivation of the kinetic equation and transport coefficients, concentrating on
interactions \cite{Kim13b} and the weak-localization
corrections \cite{Kim14}.
  A related work is also given in \cite{Burkov14},
where diffusion equations were derived to study the consequences of
chiral anomaly.
  Here we would apply a semi-classical kinetic equation,
  and the clean limit is implicitly assumed. 
  We shall 
  consider general ratios between the 
  intra and inter-valley scattering times,
  though confine ourselves to 
  relaxation time approximations. 
  Particular attention is made to the conservation of particle
  number under collisions.
  We then specialize to some specific cases where the kinetic equations
  will be solved analytically.   Our results for the electrical
  conductivities are expressed
  in terms of quantities at the Fermi level.
   Our effort here is rather closely
  related to \cite{Kim14} (but we shall ignore weak-localization
  corrections which were included there).  However, seemingly in this work
  some very special approximations have been made, limiting the applicability
  of their final results.
  We shall comment on the differences as we go along.

  We shall consider a scalar kinetic equation, that is, we limit ourselves to the case
  where the distribution
  functions are scalars, not matrices.  
  Our formulation involves directly only those states near the Fermi level.
  Hence we assume that the electronic bands are not degenerate there.
  Degeneracy at the Weyl nodes away from the Fermi level is allowed
  (and is taken care of by implicit connecting conditions  
   as done in \cite{SS13}, see also Appendix).
  For simplicity, we  shall confine ourselves to zero temperature.

  \section{The Kinetic Equation}\label{sec:KE}

  Consider then one of the bands.
  The equation of motion is given by \cite{Sundaram99}

  \bea
  {\bf \dot{r}} &=& \frac{\partial \epsilon_{\bf k}}{\partial {\bf k}}
   - {\bf \dot{k}} \times {\bf \Omega_{k}}  \label{rdot1} \\
    {\bf \dot{k}} &=&  - e {\bf E} - \frac{e}{c}
    {\bf \dot{r}} \times {\bf B} \label{kdot1}
    \eea
  where the electric charge is taken to be $-e$.  This sign convention is the same
  as \cite{Sundaram99}, but differ from those in \cite{SS13}.
  The rest of the notations are standard.  ${\bf r}$, ${\bf k}$ are
  the position and wavevector of the electron.  The band is
  specified by the dispersion $\epsilon_{\bf k}$ and curvature
  ${\bf \Omega_k}$.  ${\bf E}$ and ${\bf B}$ are the
  electric and magnetic fields.
  The solution to these equations are:
  \begin{widetext}
  \bea
   {\bf \dot{r}} &=& \frac{1}{1 + \frac{e}{c} {\bf \Omega_k \cdot B}}
   \left[ \frac{\partial \epsilon_{\bf k}}{\partial {\bf k}}
     + e {\bf E} \times {\bf \Omega_{k}} +
     \frac{e}{c} ({\bf  \frac{\partial \epsilon_{\bf k}}{\partial {\bf k}} \cdot \Omega_{k}}) {\bf B}
     \right]  \label{rdot2} \\
     {\bf \dot{k}} &=& \frac{1}{1 + \frac{e}{c} {\bf \Omega_k \cdot B}}
      \left[ - e {\bf E}  - \frac{e}{c} ({\bf  \frac{\partial \epsilon_{\bf k}}{\partial {\bf k}} \times B})
      - \frac{e^2}{c} ( {\bf E \cdot B}) {\bf \Omega_k} \right] \label{kdot2}
      \eea
      \end{widetext}
      which is also given in \cite{SS13}.  Eq (\ref{rdot2}) and (\ref{kdot2}) can also be
      directly read off from the Poisson brackets given in \cite{Duval06}.
       Note that $\frac{e}{c} {\bf \Omega_k \cdot B}$ is
      dimensionless.  Eq (\ref{rdot1}) and (\ref{kdot1}) were derived
      by expansion, order by order in gradients in ${\bf r}$ and/or ${\bf k}$.
      \cite{Panati03}
        Therefore rigorously speaking, the equations would need modifications
        near the Weyl nodes, where ${\bf \Omega_{k}}$ diverges
(e.g the phase space volume $(1 + \frac{e}{c} {\bf \Omega_k \cdot B})$ 
in eq (\ref{convtk}) and Appendix can become negative).
        We however ignore this complication and assume that the equations are valid
        also there, and derive consequences from them.
        It is also useful to note that ${\bf \Omega_k}$ and ${\bf B}$ are axial vectors,
        while ${\bf r, k, E}$ are all ordinary vectors.

        The kinetic equation then reads \cite{Xiao05}  (see also the Appendix),
        with $ n ({\bf r},{\bf k}, t)$ the occupation number,
        \be
        \frac{ \partial}{\partial t} n ({\bf r},{\bf k}, t)
        + {\bf \dot{r}} \frac{\partial}{\partial {\bf r}}  n ({\bf r},{\bf k}, t)
        + {\bf \dot{k}} \frac{\partial}{\partial {\bf k}}  n ({\bf r},{\bf k}, t)
        = I_{col}
        \label{KE0}
        \ee
        where the left hand side accounts for the drift terms in phase space, and
        the right hand side for the collisions. ${\bf \dot{r}}$ and ${\bf \dot{k}}$
        from (\ref{rdot2}) and (\ref{kdot2}) are to be used in this equation.

         We first discuss some properties
        of eq (\ref{KE0}).
        It can be checked that eq (\ref{KE0}) has the solution
        $n ({\bf r}, {\bf k}) = f (\epsilon_{\bf k} - \mu)$ in the absence
        of an electric field.  In this case, using this $n$ on the left hand side,
        the only term left is proportional to
        $ \frac{\partial \epsilon_{\bf k}}{\partial {\bf k}} \times B \cdot
        \frac{\partial n}{\partial {\bf k}}$, which is also zero due to the vector
        product.  On the other hand, the collision integral must also vanish
        for an equilibrium fermi distribution.

        The total density $n({\bf r},t)$
         at a given point in space is $\sum_{\bf k} n({\bf r}, {\bf k}, t)$.
        The total number current density ${\bf J}$ is $\sum_{\bf k} n({\bf r}, {\bf k}, t) {\bf \dot{r}}$.
        Explicitly, we have
      \begin{widetext}
        \be
        {\bf J} ({\bf r}, t) = \sum_{\bf k}
        \frac{1}{1 + \frac{e}{c} {\bf \Omega_k \cdot B}}
   \left[ \frac{\partial \epsilon_{\bf k}}{\partial {\bf k}}
     + e {\bf E} \times {\bf \Omega_{k}} +
     \frac{e}{c} ({\bf  \frac{\partial \epsilon_{\bf k}}{\partial {\bf k}} \cdot \Omega_{k}}) {\bf B}
     \right] n({\bf r}, {\bf k}, t)
     \label{J}
     \ee
     \end{widetext}
     To use the above equations, we should remember that the sum over ${\bf k}$ should be replaced
     by the integral \cite{Xiao05} (see also \cite{Duval06,Xiao06}) according to
     \be
     \sum_{\bf k} \to \int \frac{d^3 k}{(2 \pi)^3} ( 1 + \frac{e}{c} {\bf \Omega_k \cdot B})  \ .
     \label{convtk}
     \ee

   Weyl points are characterized by an integer
      $\mathfrak{M}$ referred sometimes to as the ``charge"
      of the Weyl point.  It is defined by
      \be
      \mathfrak{M} \equiv \frac{1}{2 \pi} \int d^3 k
      \frac{ \partial }{\partial {\bf k}} \cdot {\bf \Omega_k}
      \label{Mdef}
      \ee
      with the integration is over a neighborhood containing the Weyl point.
       Note that $\frac{ \partial }{\partial {\bf k}} \cdot {\bf \Omega_k}$
       vanishes everywhere except at the Weyl point, since
       ${\bf \Omega_k}$ is given by the curl of a vector potential.
      Consider the simplest example of a single Weyl node where
      the effective Hamiltonian near this point is of the form $v^* {\bf k} \cdot {\bf \tau}$ where
      ${\bf \tau}$ are $2 \times 2$ Pauli matrices and $v^* >0$.
       Then the curvature
      ${\bf \Omega_k}$, whose $z$th component evaluated via
      $\Omega_z = i \left( \langle \frac{ \partial \psi} {\partial k_y} |
      \frac{ \partial \psi} {\partial k_z} \rangle -
      \langle \frac{ \partial \psi} {\partial k_z} |
      \frac{ \partial \psi} {\partial k_y} \rangle \right) $
      where $\psi$ is the spinor wavefunction
      (and similarly for $x$ and $y$), is given by
      ${\bf \Omega_k} = \mp \frac{1}{2 k^2} \hat k$ for the positive (negative)
      energy band, where $\hat k$ the unit vector along ${\bf k}$.
      In this case, $\mathfrak{M} = \mp 1$.
      (If the effective Hamiltonian is rather $- {\bf k} \cdot {\bf \tau}$,
      then $\mathfrak{M} = \pm 1$).  Generally, $\mathfrak{M}$ must still be quantized, though
      ${\bf \Omega_k}$ does not have to have the
      isotropic form as assumed above.
      It is also helpful below to note that the value of
      $\frac{ \partial }{\partial {\bf k}} \cdot {\bf \Omega_k}$ for the
      band directly above and below the Weyl points must be opposite to each other.

      In the absence of an external electric field, the current can be shown to vanish
      (in contrast to some earlier speculations, see the Appendix, and also \cite{Zhou13,VF13}).
       To evaluate the electric current to linear order in a uniform static electric field, we can then
       subtract eq (\ref{J}) from the same expression at ${\bf E=0}$, linearizing
       in ${\bf E}$, and then multiply by the electric charge $(-e)$.
         We then obtain, for the electric current,
       \be
       {\bf J^e} = {\bf J_a^e} + {\bf J_{\rm AH}^e}
       \ee
       where
       \begin{widetext}
       \be
       {\bf J_a^e} = (-e) \int \frac{ d^3 k} {( 2 \pi)^3}
       \left[ \frac{ \partial \epsilon} {\partial {\bf k}}
        + \frac{ e}{c} (\frac{\partial \epsilon}{\partial {\bf k}} \cdot {\bf \Omega_k} )
          {\bf B} \right] \left( n ({\bf k}) - f (\epsilon_{\bf k} - \mu) \right)
        \label{Je1}
        \ee
        and
  \be
       {\bf J_{\rm AH}^e} = (-e^2) \int \frac{ d^3 k} {( 2 \pi)^3}
             {\bf E} \times {\Omega_{\bf k}} f (\epsilon_{\bf k} - \mu)
        \label{JAH}
        \ee
        \end{widetext}
        ${\bf J_{\rm AH}^e}$ involves the equilibrium distribution function
        $f (\epsilon_{\bf k} - \mu)$ and integration over all occupied levels.
        This is the familiar anomalous Hall term from the literature
        \cite{AH}.  We shall not analyze this further. It is however
        interesting to note that this term can also be written
        entirely in terms of Fermi surface quantities \cite{FS}.
         ${\bf J_a^e}$
        involves the deviation of the distribution function from the equilibrium
        value, and explicitly involves only quantities near the Fermi surface.
          We shall examine it in more detail below. Rigorously speaking
        we need to sum over all bands, but we have left out this sum to
        keep the notations simple.

        \section{Relaxation Time Approximation}\label{sec:RT}

        To obtain the conductivity tensor relating ${\bf J^e_a}$ to the uniform static
         field ${\bf E}$, we need to evaluate the distribution function
        $n ({\bf k})$ by solving the kinetic equation (\ref{KE0}) in the uniform steady state
        \begin{widetext}
        \be
       - \frac{1}{ ( 1 + \frac{e}{c} {\bf \Omega_k \cdot B}) }
      \left[  e {\bf E}  + \frac{e}{c} ({\bf  \frac{\partial \epsilon_{\bf k}}{\partial {\bf k}} \times B})
      + \frac{e^2}{c} ( {\bf E \cdot B}) {\bf \Omega_k} \right] \cdot
      \frac{ \partial n}{\partial {\bf k}}   = I_{col}
      \label{KE1}
      \ee
        In below, we shall consider two special cases and make appropriate
        approximations for the collision integral $I_{col}$.
        \end{widetext}

        \subsection{Single Fermi surface}\label{subsec:SF}
        It turns out that the Berry curvature ${\bf \Omega_k}$  has
        interesting implications which do not seem to have been fully emphasized
        in the literature.  Let us first examine this case and
        consider a single Fermi surface. ({\em There need not be
        any Weyl points for this subsection}).
        We shall assume a simple relaxation time approximation,
        with
        \be
        I_{col} = - \frac{ \delta n ({\bf k})} {\tau_0}
        \label{RT1}
        \ee
        where
        \be
        \delta n({\bf k}) \equiv n({\bf k}) - f (\epsilon_{\bf k} - \mu)
        \label{defdn}
        \ee
        Here we have already assumed that the local equilibrium distribution
        corresponds also to the chemical potential $\mu$ at equilibrium.
        This should be contrasted with subsection \ref{subsec:MP} below. Note that
        then
        $ n ({\bf k}) = f (\epsilon_{\bf k} - \mu) + \delta n({\bf k})$
         We have written down a
        momentum independent relaxation time ${\tau_0}$, though generalization
        to momentum dependent scattering rates are possible.
        In the present case, the condition of conservation of particle number in
        collision implies that
        \be
        \sum_{\bf k} \delta n ({\bf k}) = 0
        \label{mu}
        \ee
        which can be verified a posteriori.
        On the left-hand-side of eq (\ref{KE1}), there appears a term of
        the form
        $ \frac{e}{c} ({\bf  \frac{\partial \epsilon}{\partial {\bf k}} \times B})
         \cdot \frac{ \partial n}{\partial {\bf k}} $.
         [To simplify the notation, we shall occasionally leave
         out the subscript ${\bf  k}$ of $\epsilon_{\bf k}$ when
         no confusion would arise.] In the absence of $\Omega_{\bf k}$, this
         term is responsible for the classical Hall effect.  This term makes the
         solution of eq (\ref{KE1}) rather complicated for general $\Omega_{\bf k}$.
         We shall ignore this term, thus limiting ourselves to the case where
         this contribution is small compared with the ones that we are keeping.
         (See the comparison below.)

         It is convenient to define $\delta \nu (\hat k)$ via
         \be
         \delta n ({\bf k}) = - \left( \frac{\partial f}{\partial \epsilon} \right)
         \frac{ \delta \nu (\hat k)} { ( 1 + \frac{e}{c} {\bf \Omega_k \cdot B}) }
         \label{defnu}
         \ee
         Due to the factor $\left( - \frac{\partial f}{\partial \epsilon} \right)$,
         we need only to consider $\delta \nu$ on the Fermi surface, and we
         have assumed that it can be uniquely labeled by $\hat k$.
         Generalization to Fermi surfaces with more complicated shapes is
         obvious but would make the notations rather clumsy.  In this case,
         we immediately get
         \be
         \delta \nu (\hat k) = - \tau_0
            \left[  e {\bf E}
        + \frac{e^2}{c} ( {\bf E \cdot B}) {\bf \Omega_k} \right] \cdot
      \frac{ \partial \epsilon}{\partial {\bf k}}
        \label{nu1}
        \ee
        ${\bf J_a^e}$ now reads, using eq (\ref{Je1}) and (\ref{nu1})
        \begin{widetext}
       \be
       {\bf J_a^e} = e^2 \int \frac{ d^3 k} {( 2 \pi)^3}
         \left( -  \frac{ \partial f}{\partial \epsilon}   \right) \tau_0
       \left[ \frac{ \partial \epsilon} {\partial {\bf k}}
        + \frac{ e}{c} (\frac{\partial \epsilon}{\partial {\bf k}} \cdot {\bf \Omega_k} )
          {\bf B} \right]
           \left[ \frac{ \partial \epsilon} {\partial {\bf k}}
        + \frac{ e}{c} (\frac{\partial \epsilon}{\partial {\bf k}} \cdot {\bf \Omega_k} )
          {\bf B} \right] \cdot {\bf E} \
          \frac{1} { ( 1 + \frac{e}{c} {\bf \Omega_k \cdot B}) }
          \ee
          \end{widetext}

         We convert our results to integration over the Fermi surface. Using
         \bea
         \int \frac{ d^3 k}{ (2 \pi)^3} &\to&
         \int \frac{d \omega_{\hat k}}{ 4 \pi} \int \frac{ dk k^2}{ 2 \pi^2} \nonumber \\
         & \to &
         \int \frac{d \omega_{\hat k}}{ 4 \pi} \rho_{\hat k} \int d \epsilon
         \label{convtd}
         \eea
         where $\rho_{\hat k} \equiv  k^2 / [2 \pi^2 | \hat k \cdot {\bf v_F} | ]$
         is the density of states per unit solid angle $d \omega_{\hat k}$
         (here ${\bf v_F} = (\frac{\partial \epsilon}{\partial {\bf k}})_{\epsilon = \mu}$
          is the ($\hat k$ dependent) Fermi velocity), we finally get
          \begin{widetext}
          \be
        {\bf J_a^e} = e^2 \int \frac{ d \omega_{\hat k}} { 4 \pi} \rho_{\hat k} \tau_0
       \left[ \frac{ \partial \epsilon} {\partial {\bf k}}
        + \frac{ e}{c} (\frac{\partial \epsilon}{\partial {\bf k}} \cdot {\bf \Omega_k} )
          {\bf B} \right]
           \left[ \frac{ \partial \epsilon} {\partial {\bf k}}
        + \frac{ e}{c} (\frac{\partial \epsilon}{\partial {\bf k}} \cdot {\bf \Omega_k} )
          {\bf B} \right] \cdot {\bf E} /
           ( 1 + \frac{e}{c} {\bf \Omega_k \cdot B})
           \label{Je1f}
           \ee
           \end{widetext}
           An alternative way of writing (\ref{convtd}) is
           \be
         \int \frac{ d^3 k}{ (2 \pi)^3} \to
         \frac{1}{(2 \pi)^3} \int d S_{\hat k}
         \frac{1}{ |\frac{ \partial \epsilon} {\partial {\bf k}}|} \int d \epsilon \ ,
         \label{convtd2}
         \ee
         which we shall also use sometimes below.  Here $d S_{\hat k}$ denotes an
         elemental area on the Fermi surface.

           The conductivity tensor $\sigma_{ij}$ can be read off directly.
           It is simply given by
           \be
           \sigma_{a,  ij} =     e^2 \int \frac{ d \omega_{\hat k}} { 4 \pi} \tau_0 \rho_{\hat k}
              \tilde v_i  \tilde v_j /   ( 1 + \frac{e}{c} {\bf \Omega_k \cdot B})
              \label{sij}
              \ee
              where
              $\tilde v_j$ is the $j$th component of the vector
           \be
           {\bf \tilde v} \equiv       \left[ \frac{ \partial \epsilon} {\partial {\bf k}}
        + \frac{ e}{c} (\frac{\partial \epsilon}{\partial {\bf k}} \cdot {\bf \Omega_k} )
          {\bf B} \right]
          \label{vdef}
        \ee
        If ${\bf B} =0$, eq (\ref{sij}) reduces to the familiar form.
        The effect of the finite field is simply to replace
        $\frac{\partial \epsilon}{\partial {\bf k}}$ by $\bf \tilde v$,
        and the additional factor $ 1 + \frac{e}{c} {\bf \Omega_k \cdot B}$
        in the denominator.
         The subscript $a$ here reminds
         us that this is only the contribution from ${\bf J^e_a}$.  The anomalous Hall
         term is needed to complete the results. Below we shall analyze
         some special cases of eq (\ref{sij}).

         Let us consider small magnetic fields.  (\ref{sij}) can be expanded as
         \begin{widetext}
          \be
           \sigma_{a,  ij} =     e^2 \int \frac{ d \omega_{\hat k}} { 4 \pi} \tau_0 \rho_{\hat k}
           \left[ \frac{ \partial \epsilon} {\partial { k_i}}
        + \frac{ e}{c} (\frac{\partial \epsilon}{\partial {\bf k}} \cdot {\bf \Omega_k} )
          {B_i} \right]
          \left[ \frac{ \partial \epsilon} {\partial { k_j}}
        + \frac{ e}{c} (\frac{\partial \epsilon}{\partial {\bf k}} \cdot {\bf \Omega_k} )
          {B_j} \right]
           \left[ 1 - \frac{e}{c} {\bf \Omega_k \cdot B} + \frac{e^2}{c^2} ({\bf \Omega_k \cdot B})^2
           \right]
              \label{sije}
              \ee
              The term zeroth order in $B$ is
              \be
              \sigma^{(0)}_{a,  ij} =     e^2 \int \frac{ d \omega_{\hat k}} { 4 \pi} \tau_0 \rho_{\hat k}
              \frac{ \partial \epsilon} {\partial { k_i}}
                \frac{ \partial \epsilon} {\partial { k_j}}
              \ee
              which is just the ordinary expression.
              The term first order in $B$ is
              \be
              \sigma^{(1)}_{a,  ij} =     e^2 \int \frac{ d \omega_{\hat k}} { 4 \pi} \tau_0 \rho_{\hat k}
              \frac{e}{c} \left[  (\frac{\partial \epsilon}{\partial {\bf k}} \cdot {\bf \Omega_k} )
              ( \frac{ \partial \epsilon} {\partial { k_i}} B_j +
              \frac{ \partial \epsilon} {\partial { k_j}} B_i )
                -  \frac{ \partial \epsilon} {\partial { k_i}}
                \frac{ \partial \epsilon} {\partial { k_j}} {\bf \Omega_k \cdot B}
               \right]
              \ee
              This term is finite only when the system has broken time-reversal symmetry,
              as expected from general considerations.  Explicitly, if time-reversal is
              obeyed, ${\bf \Omega_{k}} = - {\bf \Omega_{-k}}$
              and $\frac{\partial \epsilon}{\partial {\bf k}}$ are odd under
              ${\bf k} \to - {\bf k}$ but the rest of the terms (excluding ${\bf B}$) are even, so
              $\sigma_{a, ij}^{(1)}$ must vanish.
              The second order terms in $B$ give
             \be
           \sigma_{a,  ij}^{(2)} =     e^2 \int \frac{ d \omega_{\hat k}} { 4 \pi} \tau_0 \rho_{\hat k}
           \left( \frac{ e}{c} \right)^2
           \left[ \frac{ \partial \epsilon} {\partial { k_i}} ({\bf \Omega_k \cdot B})
            -
          (\frac{\partial \epsilon}{\partial {\bf k}} \cdot {\bf \Omega_k} )
          {B_i} \right]
       \left[ \frac{ \partial \epsilon} {\partial { k_j}} ({\bf \Omega_k \cdot B})
            -
          (\frac{\partial \epsilon}{\partial {\bf k}} \cdot {\bf \Omega_k} )
          {B_j} \right]
              \label{sij2}
              \ee
              From this expression, it is immediately clear that each of the
              diagonal components $\sigma_{a, ii}^{(2)}$ is positive,
              giving positive magneto-conductances.  Note that our
              formulas are valid for a system with finite Berry curvatures
              ${\bf \Omega_k}$ even without Weyl nodes, so this cannot
              be assigned to axial anomalies.
              Explicitly, consider magnetic field along $\hat x$.
              Then
              \be
              \sigma_{a,  xx}^{(2)} =     e^2 \int \frac{ d \omega_{\hat k}} { 4 \pi} \tau_0 \rho_{\hat k}
           \left( \frac{ e}{c} \right)^2
              (\frac{ \partial \epsilon} {\partial { k_y}} \Omega_y
                + \frac{ \partial \epsilon} {\partial { k_z}} \Omega_z)^2 B_x^2
                \label{sxx}
                \ee
              \be
              \sigma_{a,  yy}^{(2)} =     e^2 \int \frac{ d \omega_{\hat k}} { 4 \pi} \tau_0 \rho_{\hat k}
           \left( \frac{ e}{c} \right)^2
              (\frac{ \partial \epsilon} {\partial { k_y}} \Omega_x)^2 B_x^2
              \label{syy}
                \ee
                Here we have simplified our notations and simply write
                $\Omega_x$ for the $x$ component of ${\bf \Omega_k}$ etc.
                The off-diagonal components of $\sigma_{a, ij}^{(2)}$ are
                symmetric under $i \leftrightarrow j$.
                We have
                \be
              \sigma_{a,  yx}^{(2)} =    - e^2 \int \frac{ d \omega_{\hat k}} { 4 \pi} \tau_0 \rho_{\hat k}
           \left( \frac{ e}{c} \right)^2
                  (\frac{ \partial \epsilon} {\partial { k_y}} \Omega_x)
                  \left[  (\frac{ \partial \epsilon} {\partial { k_y}} \Omega_y)
                    + (\frac{ \partial \epsilon} {\partial { k_z}} \Omega_z) \right] B_x^2
                    \ee
                    and
              \be
              \sigma_{a,  yz}^{(2)} =     e^2 \int \frac{ d \omega_{\hat k}} { 4 \pi} \tau_0 \rho_{\hat k}
           \left( \frac{ e}{c} \right)^2
                 (\frac{ \partial \epsilon} {\partial { k_y}})
                 (\frac{ \partial \epsilon} {\partial { k_z}}) \Omega_x^2 B_x^2
                 \label{syz}
            \ee

             \end{widetext}

            These terms may vanish under certain crystal symmetries.

            In the above, we have assumed one Fermi surface.
            Weyl points have never been explicitly invoked,
            and the transport properties are deduced entirely
            by consider the distribution functions near the Fermi level.
              The
            above can trivially be extended to
            the case of multiple Fermi surfaces
             provided the chemical
           potential for the Fermi surfaces can be taken to be equal.
            (see next subsection when this condition is relaxed).
            In this case we only have to sum over the contributions
            from each Fermi surface.
           Eq (\ref{sxx}) and (\ref{syy})
            implies that we can have positive magneto-conductances
            even without Weyl points.  In the present case,
            this is purely a consequence of the finite Berry curvatures.

           Our results are similar to
            those given in
            \cite{Kim14} but with some detailed differences.
            The comparison is tedious and we would
            not list them in detail here.  We comment only on some examples.
            Our (\ref{sij}) is consistent with eq (40) and (55) in \cite{Kim14}
            (when the extra contributions there (cyclotron motion and intervalley
            scattering) are dropped).  However, they did not reach
            our (\ref{sxx})-(\ref{syz}).  For example,
            their (57), (42),
            when specialized to our present situation, differ from
            our (\ref{sxx}) and (\ref{syy}).
            Seemingly additional implicit assumptions were made
            in \cite{Kim14} in the derivations of their subsequent equations.

            Let us estimate the order of magnitude of the magneto-conductance,
            assuming that the term first order in $B$ vanishes.
            From the above, we see that the ratio of
            the conductivity second order in $B$ relative to
            the zeroth order (for any direction $j$) can be estimated as, e.g.,
            \be
            \frac{\sigma^{(2)}_{a,jj}}{\sigma^{(0)}_{a,jj}}
            \sim \left(\frac{e}{\hbar c} \tilde \Omega B \right) ^2
            \label{est}
            \ee
            where $\tilde \Omega$ is a typical value of ${\bf \Omega_k}$ on
            the Fermi surface.  Here
            we have also restored the $\hbar$ which was set to unity above.
            A convenient form of this dimensionless ratio can also be written as
             $ (\tilde \Omega / l_B^2 )^2$, where
             $l_B \equiv ( \hbar c / e B)^{1/2}$ is the magnetic length.
             Numerically, (it is convenient to note that
             $\pi \hbar c / e = 2.07 \times 10^{-7}$ in c.g.s. units)
             we find that, with
             $\Delta \sigma = \sigma(B) - \sigma(0)$ being the
             change in conductivity by $B$ (we now leave out the
             subscripts for simplicity),
             \be
             \frac{\Delta \sigma}{\sigma(0)}
             \approx 1.79 \times 10^{-11}
             \left( \frac{\tilde \Omega}{\rm Bohr^2} \right)^2
             \left( \frac{B}{\rm Testla} \right)^2
             \label{est2}
             \ee

             This magneto-conductance is significant only when
             $\tilde \Omega$ is sufficiently large.
          To have
             change in conductivity of say   $~ 2 \%$
              in a field of one Testla,
              $\tilde \Omega$ has to be larger than $~ 3 \times 10^4/{\rm Bohr}^2$.

             In the above, we have ignored the cyclotron motion of
             the electrons (see discussion below  eq (\ref{mu}))
             which typically gives a positive magneto-resistance
             for current flows perpendicular to the magnetic field.
             This contribution can be estimated by
             \be
             \frac{\Delta \sigma_{\rm cy}}{\sigma(0)}  \sim - (\omega_c \tau)^2
             \ee
             where $\omega_c$ is the cyclotron frequency.
             The right-hand-side of the above can also be written
             as $ (l / k_F l_B^2)^2$ where $l$ is the mean-free path.
             For our magneto-conductance to be more significant than
             this magneto-resistance contribution requires
             $k_F \tilde \Omega > l$.

         \subsection{Multiple Fermi surfaces}\label{subsec:MP}

           In this section, we consider multiple Fermi surfaces.
           This situation may include the one where
           each Fermi surface contains one (or multiple) Weyl point with finite net charge.
           A particular limit of this situation has already been considered
           in \cite{SS13}, assuming that the intra-Fermi surface scattering
           is infinitely fast compared with inter-Fermi surfaces scattering.
           (There are additional implicit assumptions taken which we shall discuss
           further below).  Our main aim here is to relax this (and the other)
           assumption(s).  We shall also indicate how the results of Ref \cite{SS13}
           can be recovered as a special limit.
           (We note however that the formulas below would also be applicable
           to the case where there are simply two Fermi pockets without
           Weyl nodes, though the resulting magneto-conductance is expected
           to be much smaller: see below).

           For definiteness, let us consider two Fermi pockets, labeled by
           $1$ and $2$, though the generalization to multiple pockets is
           straight-forward with the expense of rapid increase in
           the number of indices and the number of self-consistent
           equations for the chemical potentials (see (\ref{mud}) and (\ref{mus}) below).
           Let us first consider
           pocket $1$.  For the intra-pocket scattering,
           we can write a form analogous to  eq (\ref{RT1}):
           \be
           I_{col}^{(11)} = - \frac{\delta n ({\bf k})}{\tau_0}
           \label{intra}
           \ee
           The superscript in eq (\ref{intra}) denotes that we are considering
           scattering from pocket $1$ to $1$.
           \be
           \delta n ({\bf k}) = n ({\bf k}) - f (\epsilon - \mu_1)
           \label{le}
           \ee
           now denotes the deviation from local equilibrium.
           Eq (\ref{le}) demands
           that the relaxation is towards local equilibrium for this
           Fermi pocket towards chemical potential $\mu_1$, which
           is to be determined later. The condition that particle
           numbers are conserved under intra-pocket scattering implies
           \be
           \sum_{{\bf k} \in 1} \delta n ({\bf k}) = 0
           \label{mu1}
           \ee
           similar to eq (\ref{mu}),
           though in the previous section we can directly take $\mu_1$ to
           be the equilibrium chemical potential.

           There are also inter-pocket scattering.
           To write down a possible term for the collision integral, it is important here to
           pay special attention to conservation laws \cite{PN}, in particular
           for particle number. For pocket $1$, let us try
           \cite{footnoteKim14}
           \be
           I^{(12)}_{col} =
           (n ({\bf k}) - f (\epsilon - \mu_2))/\tau_k^{12} \ ,
           \label{I12}
           \ee
           where $\mu_2$ is
           the chemical potential of pocket $2$ and $\tau_k^{12}$ a relaxation
           time, with a similar
           term for pocket $2$.  $\mu_{1}$ is to be determined
           by (\ref{mu1}), and a similar equation applies to
           $\mu_2$.  The conservation of particle number
           demands that, for any distribution $n ({\bf k})$, we must have
           \be
           \sum_{{\bf k} \in 1}
           \frac{n ({\bf k}) - f (\epsilon - \mu_2)}{\tau_k^{12}}
           +
           \sum_{{\bf k} \in 2}
           \frac{n ({\bf k}) - f (\epsilon - \mu_1)}{\tau_k^{21}} = 0 \ ,
           \label{conser1}
           \ee
           This places restrictions on the scattering times $\tau_k^{12}$
           and $\tau_k^{21}$.  One possible choice is that
           both these quantities are independent of $k$.  Eq (\ref{conser1})
           then becomes
           \be
           \sum_{{\bf k} \in 1}
           \frac{f (\epsilon - \mu_1) - f (\epsilon - \mu_2)}{\tau^{12}}
           +
           \sum_{{\bf k} \in 2}
           \frac{f (\epsilon - \mu_2) - f (\epsilon - \mu_1)}{\tau^{21}} = 0
           \label{conser2}
           \ee
           Linearizing in $\mu_1 - \mu_2$ and using (\ref{convtk}) and (\ref{convtd}),
           we see that
           this is obeyed if
           \be
           \frac{\tilde D_1 (B)}{\tau^{12}}
           = \frac{\tilde D_2 (B)}{\tau^{21}}
           \label{conser3}
           \ee
           where we have defined
           \be
            \tilde D_1 (B) \equiv
            \int_{k \in 1} \frac{d \omega_{\hat k}}{ 4 \pi} \rho_{\hat k}
              (1 + \frac{e}{c} {\bf \Omega_k \cdot B})
              \label{tD}
              \ee
             ( with the integral over the Fermi surface $1$)
              and similarly for $\tilde D_2$.
              It is convenient to define $\tau^X$ so that
           \be
           \frac{ 1}{\tau^{12}} = \frac{1}{\tau^X}
           \frac{\sqrt{\mathfrak{D}_1 \mathfrak{D}_2}}{\tilde D_1 (B) }
           \ee
           and also with $ 1 \leftrightarrow 2$.
           Here
           $D_1 \equiv \int_{{\bf k} \in 1} \frac{d \omega_{\hat k}}{ 4 \pi} \rho_{\hat k}$
            is the density of states for the Fermi pocket $1$ in zero field.
            It is feasible to take $\tau^X$ as field independent, but
            then $\tau^{12}$ and $\tau^{21}$ must depend on $B$.

            Using
                    \be
           n ({\bf k}) =
                 [n ({\bf k}) - f(\epsilon - \mu_1)] + f(\epsilon - \mu_1)
           \label{nknu}
           \ee
            we get
            \be
            I^{(12)} = -\frac{1}{\tau^X}
             \frac{\sqrt{\mathfrak{D}_1 \mathfrak{D}_2}}{\tilde D_1 }
             \left[ \delta n ({\bf k}) - \frac{\partial f}{\partial \epsilon}
               (\mu_1 - \mu_2) \right]
               \label{inter}
             \ee
             This form makes clear that the sum $\sum_{k \in 1} I^{(12)}$
              is finite only if $\mu_1 \ne \mu_2$, and represents
              the decrease (increase) in number of particles for Fermi pocket 1
              if $\mu_1 > (<) \mu_2$ via intervalley collisions.  By construction, the change
              in total particle numbers of the two Fermi pockets via collisions
              must be zero.

             (\ref{inter}), together with the intra-pocket contribution (\ref{intra}),
             gives us in total the collision terms
           \be
           I_{col}^1 = - \frac{ \delta n ({\bf k})}{ \tilde \tau_1}
           + \frac{\partial f}{\partial \epsilon}
           \frac{1}{\tau^X}
             \frac{\sqrt{\mathfrak{D}_1 \mathfrak{D}_2}}{\tilde D_1 }
            (\mu_1- \mu_2)
             \label{Icol1} \ee
             where we have defined
             \be
             \frac{1}{\tilde \tau_1} = \frac{1}{\tau_0}
               + \frac{1}{\tau^X}
             \frac{\sqrt{\mathfrak{D}_1 \mathfrak{D}_2}}{\tilde D_1 (B) }
             \ee
             $\tilde \tau_1$ is in general $B$ dependent.
            Defining $\delta \nu (\hat k)$ as in eq (\ref{defnu})
            (note however that $\delta n( {\bf k})$ is now the deviation
            from local equilibrium, {\it i.e.}, eq (\ref{le})),
            we have
            \begin{widetext}
            \be
           I_{col}^1 = - \frac{ \partial f} {\partial \epsilon}
           \frac{1} { ( 1 + \frac{e}{c} {\bf \Omega_k \cdot B}) }
           \left(
           - \frac{ \delta \nu (\hat k)}{\tilde \tau}
           - \frac{ \mu_1 - \mu_2} {\tau^X}
           \frac{\sqrt{\mathfrak{D}_1 \mathfrak{D}_2} }{\tilde D_1}
           ( 1 + \frac{e}{c} {\bf \Omega_k \cdot B})
            \right)
           \label{I1f}
              \ee
              With this, eq (\ref{KE1}) becomes,
              after dropping the term
           involving ${\bf  \frac{\partial \epsilon}{\partial {\bf k}} \times B}$
           under the assumption as stated in the last section
           and linearizing with respect to the electric field,
%
%
           \be
              \left[  e {\bf E}
      + \frac{e^2}{c} ( {\bf E \cdot B}) {\bf \Omega_k} \right] \cdot
      \frac{ \partial \epsilon}{\partial {\bf k}}   =
             - \frac{ \delta \nu (\hat k)}{\tilde \tau}
            - \frac{ \delta \mu_1 - \delta \mu_2} {\tau^X}
                               \frac{\sqrt{\mathfrak{D}_1 \mathfrak{D}_2} }{\tilde D_1}
           ( 1 + \frac{e}{c} {\bf \Omega_k \cdot B})
             \,
            \label{KEf}
           \ee
            where we have defined $\delta \mu_{1,2} \equiv \mu_{1,2} - \mu$ with $\mu$ the
            equilibrium chemical potential.  (If we have more bands,
            then more terms proportional to chemical potential differences
            would appear on the right hand side of (\ref{KEf})).

           For ${\bf k} \in 1$, using (\ref{nknu})
            in eq (\ref{Je1}) and using
               eq (\ref{defnu}), we obtain the contribution to ${\bf J_a^e}$ from Fermi pocket $1$

          \be
        {\bf J_{a1}^e} = - e \int_{{\bf k} \in 1} \frac{ d \omega_{\hat k}} { 4 \pi} \rho_{\hat k}
       \left[ \frac{ \partial \epsilon} {\partial {\bf k}}
        + \frac{ e}{c} (\frac{\partial \epsilon}{\partial {\bf k}} \cdot {\bf \Omega_k} )
          {\bf B} \right]
           \left[ \frac{ \delta \nu (\hat k)} { ( 1 + \frac{e}{c} {\bf \Omega_k \cdot B})}
             + \delta \mu_1 \right]
           \label{Je2}
           \ee

           It remains to solve for $\delta \nu (\hat k)$ and $\delta \mu_{1,2}$.
           The latter are determined by conservation of particle number.
            Operating $\int_{{\bf k} \in 1} \frac{ d \omega_{\hat k}}{ 4 \pi}  \rho_{\hat k}$ on
           eq (\ref{KEf}) and using (\ref{mu1}) (which gives
           $\int_{{\bf k} \in 1} \frac{ d \omega_{\hat k}}{ 4 \pi}  \rho_{\hat k} \delta \nu (\vec k) = 0$),
           we get
           \be
           \delta \mu_1 - \delta \mu_2
            =  - \tau^X \ \frac{ \int_{{\bf k} \in 1}  \frac{ d \omega_{\hat k}}{ 4 \pi}  \rho_{\hat k}
                    \left[  e {\bf E} \cdot
      \frac{ \partial \epsilon}{\partial {\bf k}}
      + \frac{e^2}{c} ( {\bf E \cdot B}) {\bf \Omega_k}  \cdot
      \frac{ \partial \epsilon}{\partial {\bf k}} \right] }
        { (D_1 D_2)^{1/2} }
        \label{mud}
        \ee
       \end{widetext}
         Note that we have a similar equation with $1 \leftrightarrow 2$.
          These two equations are consistent with each other only
          if the numerators sum to zero, that is, alternatively,
          the integral of the left hand side of (\ref{KEf}) over ${\bf k}$
          vanishes when also summed over all bands.
          See the Appendix for further discussions.

        On the other hand, using (\ref{nknu}) and a similar equation for
        pocket $2$, the requirement that the total particle number be the
        same as equilibrium requires, with again the help of eq (\ref{mu1}),
        \be
         \tilde D_1 \ \delta \mu_1 \
           + \tilde D_2 \ \delta \mu_2 \ = 0
        \label{mus}
           \ee
        Eq (\ref{mud}) and (\ref{mus}) then determines $\delta \mu_{1,2}$.  Once
        they are determined, $\delta \nu (\hat k)$ can then be obtained from
        eq (\ref{KEf}) and the current from (\ref{Je2}).

        Let us first consider the special case where
        the two Fermi pockets are located at $\pm {\bf k_0}$ and related by
        inversion symmetry.  In this case, if ${\bf k}\in 1$ lies near ${\bf k_0}$, then
        $-{\bf k} \in 2$ lies near $-{\bf k_0}$, with ${\bf \Omega (-k)} = {\bf \Omega (k)}$
        but $\frac{ \partial \epsilon}{\partial {\bf k}} |_{- \bf k}
         = - \frac{ \partial \epsilon}{\partial {\bf k}} |_{\bf k}$.
         Note that these two Fermi pockets together must enclose zero total charge $M$.
         Eq (\ref{mus}) then implies that $\delta \mu_1 = - \delta \mu_2$,
         as expected, and so the left hand side of (\ref{mud}) simply
         becomes $2 \delta \mu_1$, with furthermore
        $D_1 = D_2$ (and $\tilde D_1 = \tilde D_2$) which we shall simply denoted as $D$
        ($\tilde D$).
         Eq (\ref{KEf}) then gives
              \begin{widetext}
         \be
         \delta \nu (\hat k) = - \tilde \tau_1
         \left[  e {\bf E}
        + \frac{e^2}{c} ( {\bf E \cdot B}) {\bf \Omega_k} \right] \cdot
      \frac{ \partial \epsilon}{\partial {\bf k}}
       - 2 \frac{\tilde \tau_1}{\tau^X}
       \frac{\mathfrak{D}}{\tilde D}
           ( 1 + \frac{e}{c} {\bf \Omega_k \cdot B})
       \delta \mu_1
       \ee
       Inserting this into eq (\ref{Je2}) gives us finally,
       using
       the short-hand $\tilde {\bf v}$ defined in eq (\ref{vdef})
        for
       the contribution {\em from pocket 1},

       \bea
       {\bf J^e_{a1}} &=&
       e^2 \tilde \tau_1 \int_{k \in  1} \frac{ d \omega_{\hat k}}{ 4 \pi}  \rho_{\hat k}
        \frac{ \tilde {\bf v} [ \tilde {\bf v} \cdot {\bf E} ]}
          { ( 1 + \frac{e}{c} {\bf \Omega_k \cdot B}) } \nonumber \\
           & & + \frac{ e^2 \tau^X}{2 D}
           \left( 1 - 2 \frac{\tilde \tau_1}{ \tau^X}
            \frac{D}{\tilde D} \right)
          \left[ \int_{k \in 1} \frac{ d \omega_{\hat k}}{ 4 \pi}  \rho_{\hat k} \tilde {\bf v}
             \right]
            \left[ \int_{k \in 1} \frac{ d \omega_{\hat k}}{ 4 \pi}  \rho_{\hat k} \tilde {\bf v} \cdot {\bf E}
            \right]
            \label{Je2f}
            \eea
         Pocket $2$ gives an equal contribution, so that
          the total ${\bf J^e_{a}}$ is just twice of ${\bf J^e_{a1}}$.
          The first term of (\ref{Je2f}) is the same as what had in the last section
          (eq (\ref{Je1f})), except that $\tau_0 \to \tilde \tau_1$,
          which is in general also field dependent.   The second term is a
          new contribution.

          \end{widetext}

           The zero field conductances and finite field magneto-conductances
           can be obtained by expanding eq (\ref{Je2f}).
           (recall that $\tilde \tau_1$, $\tilde D$, ${\bf \tilde v}$
           are all field dependent). The resulting
           formulas for the general situation are quite lengthy.
          We simply indicate here the relations between our results
          and that in Ref \cite{SS13}.
          Our expression for the current generalizes that of \cite{SS13} in that we have
          also included the effects of deviation from local equilibrium $\delta \nu (\hat k)$,
          which gives the first term in eq (\ref{Je2f}) (and also a correction
          to the second term).
          This contribution can be dropped if $\tau_0 << \tau^X$.
          With this approximation, we now have
               \be
       {\bf J^e_{a1}} =
         \frac{ e^2 \tau^X}{2 D}
          \left[ \int_{k \in 1} \frac{ d \omega_{\hat k}}{ 4 \pi}  \rho_{\hat k} \tilde {\bf v}
            \right]
            \left[ \int_{k \in 1} \frac{ d \omega_{\hat k}}{ 4 \pi}  \rho_{\hat k} \tilde {\bf v} \cdot {\bf E}
            \right]
            \label{Je2ff}
            \ee
           In this equation, $ \int_{k \in 1} \frac{ d \omega_{\hat k}}{ 4 \pi}  \rho_{\hat k}
          \frac{ \partial \epsilon} {\partial {\bf k}}$ in general vanishes,
          leaving us only a second order term in ${\bf B}$.
          Defining
           \be
           \int_{k \in 1} \frac{ d \omega_{\hat k}}{ 4 \pi}  \rho_{\hat k}
          \frac{ \partial \epsilon} {\partial {\bf k}} \cdot \Omega ({\bf k})
           = \frac{\tilde M_1}{ 4 \pi^2}
           \label{tM}
           \ee
           we get, for the total current from both pockets,
           (with $\tilde M_1 = \tilde M_2 = \tilde M$ in the present case)
           \be
           {\bf J^{e,(2)}_{a,tot}} =
           \tau^X \frac{e^4 \tilde M^2}{  ( 4 \pi^2)^2 c^2 D} {\bf B} ( {\bf B} \cdot {\bf E})
           \label{JSS}
           \ee
           The magneto-conductance depends on $\tilde M$, an integral
           on the Fermi surface, as one expects on general grounds for a transport property.
           Using (\ref{convtd}) and (\ref{convtd2}), we have
          \be
           \frac{\tilde M_1}{ 4 \pi^2}
           = \frac{1}{ (2 \pi)^3} \int_{k \in 1} d S_{\hat k}
           {\hat {\bf n}} \cdot {\bf \Omega (k)}
           \ee
           where $\hat {\bf n} \equiv \frac{\partial \epsilon}{\partial {\bf k}}
           / |\frac{\partial \epsilon}{\partial {\bf k}}|$ is the
           unit vector normal to the Fermi surface.
           Employing Gauss theorem, $\tilde M_1$ is then seen to be proportional to
           the total outward flux of $\bf \Omega (k)$ from the Fermi surface.
           Since $\bf \Omega(k)$ is divergence free, $\tilde M_1$ is thus quantitized. 
           If there is a single Weyl point inside the Fermi surface,
           $\tilde M$ is just $M$, the charge of the Weyl point it encloses.
             More generally, it is a sum over the charges.
            If one assumes that only one isotropic Weyl point of unit charge is enclosed, using
           $D = k_F^2/ (2 \pi^2 v_F) = \mu^2 / 2 \pi^2 v_F^3$ with $k_F$ the Fermi momentum, then
           eq (\ref{JSS}) reduces to that of \cite{SS13} except with a difference in notations.
           ($\tau^X/2 \to \tau$ there: see eq (\ref{I12})).

           Let us return to the more general case where the two Fermi pockets
           are not related by symmetry.  Solving for $\delta \mu_{1,2}$ from
           eq (\ref{mud}) and (\ref{mus}) and substituting in eq (\ref{Je2}),
           we find that the current ${\bf J_{a1}^e}$ is given by
           the same expression as in (\ref{Je2f}), provided we use the replacement
           \begin{widetext}
           \bdm
           \frac{ 1}{2 D}
           \left( 1 - 2 \frac{\tilde \tau_1}{ \tau^X}
            \frac{D}{\tilde D} \right) \to
            \frac{1}{(D_1 D_2)^{1/2}}
            \frac{\tilde D_2}{ \tilde D_1 + \tilde D_2}
                   \left( 1 -  \frac{\tilde \tau_1}{ \tau^X}
            \frac{(D_1 D_2)^{1/2} (\tilde D_1 + \tilde D_2)}{\tilde D_1 \tilde D_2} \right)
            \edm
            \end{widetext}
            in the second term.   In the $\tau_0 \ll \tau^X$ limit,
            eq (\ref{JSS}) holds with the replacement
            $D \to (D_1 D_2)^{1/2}$ in the denominator.

           Despite the more involved form of (\ref{JSS}), the order of magnitude for
          the magneto-conductance can still be estimated in similar manner
           as in subsection \ref{subsec:SF}.
           Since the zero field conductivity is dictated rather by the first
           term in (\ref{Je2f}), there is an extra factor
           $\sim \tau^X/\tau_0$ for the ratio $\Delta \sigma / \sigma(0)$
           in eq (\ref{est2}).  If the Fermi pockets do not enclose any
           Weyl points, $\tilde M = 0$ and the second term of eq (\ref{Je2f}) vanishes,
           the estimate of the last section applies
            giving much smaller magneto-conductances.

          \section{Summary}
          In this paper, employing a semi-classical kinetic equation with
          relaxation time approximation and taking into
          account the effect of Berry curvature in Weyl metals,
           we evaluate the magneto-conductivity at low fields in the clean limit.
          We pay special attention to conservation of particles
          when writing down the collision integrals.  We showed
          how our results reduce to those \cite{SS13} in a special limit.
          We hope that our more generalized equations can provide
          better understanding of transport in Weyl systems,
          especially when system parameters are known in more detail.

\section{Acknowledgement}
This work is supported by the Ministry of Science and Education in
Taiwan under grant number MOST-101-2112-M-001-021-MY3.
I also thank my colleagues in the nano-science group for
motivating me to perform this work and many discussions.

\appendix

\section{Further discussions on the kinetic equation }\label{app1}

  \paragraph{ }
    Let us denote the phase space volume by $\mathcal{R} ({\bf r}, {\bf k}, t)$.  In our case,
    $ \mathcal{R}= 1 + \frac{e}{c} {\bf \Omega_k}
    \cdot {\bf B}$.   On general grounds, one expects
    $\mathcal{R}$ to obey a continuity equation in phase space
    \be
    \frac{\partial \mathcal{R}}{\partial t} +
    \frac{\partial}{\partial {\bf r}} \cdot ( \dot{\bf r} \mathcal{R}) +
    \frac{\partial}{\partial {\bf k}} \cdot ( \dot{\bf k} \mathcal{R})  = 0
    \label{contD}
    \ee
    This equation has also been given in \cite{Xiao05} though in a slightly different
    form.  The kinetic equation, in the absence of collisions, can also be written as
    \be
    \frac{\partial ( n \mathcal{R} ) }{\partial t} +
    \frac{\partial}{\partial {\bf r}} \cdot ( \dot{\bf r} n \mathcal{R}) +
    \frac{\partial}{\partial {\bf k}} \cdot ( \dot{\bf k} n \mathcal{R})  = 0
    \label{contnD}
    \ee
    The continuity equation (\ref{KE1}) follows from using (\ref{contD}) in (\ref{contnD})
    (after adding the collision contribution).

    Let us check examine (\ref{contD}) directly by using
    (\ref{rdot2}) and (\ref{kdot2}), in the case where
    the material parameters such as $\bf \Omega_k$ are time and
    space independent.    We have
    \bdm
     \frac{\partial \mathcal{R}}{\partial t} = \frac{e}{c}
     \frac{\partial {\bf B}}{\partial t} \cdot {\bf \Omega_k}
     \edm
     \bdm
        \frac{\partial}{\partial {\bf r}} \cdot ( \dot{\bf r} \mathcal{R})
         = e {\bf \Omega_k} \cdot ({\bf \nabla} \times {\bf E})
         + \frac{e}{c} (\frac{\partial \epsilon}{\partial {\bf k}} \cdot {\bf \Omega_k} )
         {\bf \nabla} \cdot {\bf B}
         \edm
         and
         \bdm
         \frac{\partial}{\partial {\bf k}} \cdot ( \dot{\bf k} \mathcal{R})
          = \frac{e^2}{c} ( {\bf E} \cdot {\bf B}) ({\bf \nabla_k} \cdot {\bf \Omega_k})
          \edm

         For a general point ${\bf k}$, since the curvature ${\bf \Omega_k}$ is
      given by the curl of a vector potential,
      $\frac{ \partial }{\partial {\bf k}} \cdot {\bf \Omega_k}$
       vanishes.  Hence,
       in the absence of Weyl points, eq (\ref{contD}) is indeed valid
          after we invoke the Maxwell equations
          ${\bf \nabla} \cdot {\bf B} = 0$ and
          $\frac{1}{c}\frac{\partial {\bf B}}{\partial t}
          + {\bf \nabla} \times {\bf E} = 0$.
          With a Weyl point, there is an additional source term
            $\frac{ \partial }{\partial {\bf k}} \cdot {\bf \Omega_k}$.

       This has some unusual consequences.
      The continuity equation has to be replaced by \cite{SS13}
      one with an extra source term $\mathcal{S}$, {\it i.e.},
      \be
      \frac{\partial n({\bf r},t)}{\partial t}
       + {\bf \nabla} \cdot J ({\bf r}, t) - \mathcal{S} = \sum_k I_{col}
       \label{cont}
       \ee
       where
       \be
       \mathcal{S} = - \frac{e^2}{ 4 \pi^2 c} ( {\bf E} \cdot {\bf B}) f(\epsilon^* - \mu)
        \mathfrak{M}
        \label{source}
        \ee
            The right hand side of (\ref{cont}) denotes the contribution to the
       change in density from collision, but the term $\mathcal{S}$ on the left
       denotes the contribution entirely from the drift in phase space.

       This source term in the continuity equation
       has the same origin as the chiral anomaly discussed
       in \cite{NM83} for strong magnetic field.  In our case, it
        can be understood by examining (\ref{kdot2}).
       To simplify our discussion, let us pretend that the Weyl point
       at ${\bf k = 0}$ and the dispersion is isotropic around this point.
        We then have $\bf \Omega_k = \frac{ \mathfrak{M} \hat k}{2 k^2}$.
        For the band  with energy above the Weyl point,
        $\epsilon_{\bf k} = \epsilon^* + v^* k$ for some $v^* > 0$.
        Then we have for ${\bf \dot{k}}$, up to the factor
       $1 + \frac{e}{c} {\bf \Omega_k \cdot B}$, the term
        $- \frac{e^2}{c} ( {\bf E \cdot B}) {\bf \Omega_k}
         = - \frac{e^2}{c} ( {\bf E \cdot B}) \frac{ \mathfrak{M} \hat k}{2 k^2}$.
         If $({\bf E \cdot B}) \mathfrak{M} < (>) 0$, then we have particles coming
         from (disappearing to) the origin and going towards finite $k$. The total number is
         precisely given by the term $\mathcal{S}$.  The presence of this
         term should not alarm us:  for the band right below, $\mathfrak{M}$ must
         be exactly opposite, meaning simply that the additional particles from
         one band must come from the other through this Weyl point.

         Alternatively, one can consider the change in the kinetic energy of a particle
         due to the drift terms.  This is given by ${\bf \dot{k}}
         \cdot \frac{\partial \epsilon_k}{\partial {\bf k}}$.  Besides the
         ordinary terms proportional to ${\bf E} \cdot  \frac{\partial \epsilon_k}{\partial {\bf k}}$
         describing the work done by the electric field, we have
         the additional term
         $- \frac{e^2}{c} ( {\bf E \cdot B})
         {\bf \Omega_k} \cdot  \frac{\partial \epsilon_k}{\partial {\bf k}}$
         Hence, if ${\bf E \cdot B} < 0$ say, then the kinetic energy is increasing if
          ${\bf \Omega_k}$ and the velocity $\frac{\partial \epsilon_k}{\partial {\bf k}}$
          are antiparallel, showing that the energy of the particle would increase with time.
          If the particle is originally below and near the Weyl point,  it would eventually
          goes through the Weyl point and emerges above it.

          For $\mathcal{R}$, this source term can be understood in the same manner
          as the source term for extra particles discussed two paragraphs above.
          We note however that our derivation of magneto-conductance in text never invoked
          explicitly the continuity equation (\ref{cont}).  However,
          the difference between ($n$ times) (\ref{contD}) and (\ref{contnD})
          yields the term proportional to
          $ ({\bf E \cdot B}) {\bf \Omega_k} \cdot  \frac{\partial n}{\partial {\bf k}}$
          in, e.g., (\ref{KE1}), and eventually responsible for the appearance of
          $\tilde M$ in eq (\ref{tM}) and the mangeto-conductance (\ref{JSS}).

  \paragraph{ }
         We next show here that, (i) in the absence of an external electric field
        the electric current vanishes, and (ii), the eq (\ref{mud}) is
        consistent with itself with $1 \leftrightarrow 2$.
          Equation (\ref{J}) gives, with $n({\bf k})$ given by the equilibrium Fermi function
          $f(\epsilon_{\bf k} - \mu)$, the expression
            \be
        {\bf J}  = \sum \int \frac{d^3 k}{ (2 \pi)^3}
   \left[ \frac{\partial \epsilon_{\bf k}}{\partial {\bf k}} +
     \frac{e}{c} ({\bf  \frac{\partial \epsilon_{\mathbf k}}{\partial {\bf k}} \cdot \Omega_{k}}) {\bf B}
     \right] f(\epsilon_{\bf k} - \mu)
     \label{J0}
     \ee
     with a sum over all bands,
     where these bands need not intersect the Fermi level.
     We note that the right hand side of (\ref{mud}) involves
     a very similar integral as in eq (\ref{J0}).
     Let us define
     \be
     {\bf \mathfrak{F}}_1 (\epsilon')
     =  \int_{k \in 1} \frac{d^3 k}{ (2 \pi)^3}
   \left[ \frac{\partial \epsilon_{\bf k}}{\partial {\bf k}} +
     \frac{e}{c} ({\bf  \frac{\partial \epsilon_{\mathbf k}}{\partial {\bf k}} \cdot \Omega_{k}}) {\bf B}
     \right] \delta (\epsilon - \epsilon')
     \label{dF1}
     \ee
      Eq (\ref{mud}) is then
     \be
     \delta \mu_1 - \delta \mu_2 = - \tau^X \frac{
     {\bf \mathfrak{F}}_1 (\mu) \cdot {\bf E}}
     { (D_1 D_2)^{1/2}}
     \ee
     and the equilibrium current is
     \be
     {\bf J} = \int d \epsilon' {\bf \mathfrak{F}}_{\rm tot}  (\epsilon')
      f (\epsilon' - \mu)
     \ee
     where ${\bf \mathfrak{F}}_{\rm tot}$ is the sum  $\sum {\bf \mathfrak{F}}_j $ over bands
     $j$.

     It is then sufficient to prove that the sum ${\bf \mathfrak{F}}_{\rm tot} (\epsilon')$
      vanishes
     for any $\epsilon'$.
     Let us call the first and second contributions in eq (\ref{dF1})
     ${\bf \mathfrak{F}}^a_{1}$ and ${\bf \mathfrak{F}}^b_{1}$, with
     similar definitions for ${\bf \mathfrak{F}}^{a,b}_{\rm tot}$.
     The vanishing of ${\bf \mathfrak{F}}^a_{\rm tot}$ is easiest to see.
     Using (\ref{convtd2}), we see that the $x_i$ component of
     ${\bf \mathfrak{F}}^a_{\rm tot}$ is proportional to the
     the surface integral of the dot product of $\hat x_i$ with the
     Fermi surface normal ${\hat {\bf n}}$
     ( $\equiv \frac{\partial \epsilon}{\partial {\bf k}}
           / |\frac{\partial \epsilon}{\partial {\bf k}}| $) and hence vanishes
     since the Fermi surfaces must form close surfaces.

      Let us now consider ${\bf \mathfrak{F}}^b_{\rm tot}$.
     The contribution from band 1 is proportional to
     $\int_{{\bf k} \in 1}  \frac{ d \omega_{\hat k}}{ 4 \pi}  \rho_{\hat k}
                    {\bf \Omega_k}  \cdot
      \frac{ \partial \epsilon}{\partial {\bf k}}$,
      with the integral now over a surface at energy $\epsilon'$.
      We show that, when summed over all bands,
      this quantity vanishes.  One way to proceed is to note
      that the above integral is proportional to $\tilde M_1$  defined in (\ref{tM})),
      and then uses the Fermion doubling theorem \cite{NM83}.
      However, we may also proceed as follows.
%
%
      We write this sum, using the inverse of eq (\ref{convtd}), as
     $- \sum \int \frac{d^3k}{(2 \pi)^3} {\bf \Omega_k}
     \cdot \frac{\partial n}{\partial {\bf k}}$.
     where the integral is now over all momentum ${\bf k}$, and the sum is over all bands,
     and  $n = f(\epsilon - \epsilon')$.
     Using integration by parts and periodicity of
     $n ({\bf k}) \bf {\Omega_{k}}$ over the Brillouin zone,
     this sum can be rewritten as
     \bdm
     \sum \int \frac{d^3k}{(2 \pi)^3} n
      \frac{\partial}{\partial {\bf k}} \cdot {\bf \Omega_k } \ .
      \edm
     It is obviously zero when all bands are empty.
      Its value can change only when a point where
      $\frac{\partial}{\partial {\bf k}} \cdot {\bf \Omega_k } \ne 0$ changes its
      occupation, but then the contribution above and below this Weyl point
      cancels, since
      we can replace $n$ by unity and the bands above and below the
      Weyl points have opposite $ \frac{\partial}{\partial {\bf k}} \cdot {\bf \Omega_k }$.
      [It is may be helpful also to put the argument slightly differently:
      (a)  the sum of   $\int_{{\bf k}}  \frac{ d \omega_{\hat k}}{ 4 \pi}  \rho_{\hat k}
                    {\bf \Omega_k}  \cdot
      \frac{ \partial \epsilon}{\partial {\bf k}}$ over the bands
      vanish when $\epsilon'$ is below all Weyl points
      using the integration by parts argument given above;
      (b) This sum does not change with $\epsilon'$ when the Weyl points
      are not crossed (also via the arguments above); and finally
      (c) when $\epsilon'$ increases from slightly below to slightly above a Weyl point,
      such that say band 1 changes from a small hole pocket to an electron pocket,
      the integral
          $\int_{{\bf k} \in 1}  \frac{ d \omega_{\hat k}}{ 4 \pi}  \rho_{\hat k}
                    {\bf \Omega_k}  \cdot
      \frac{ \partial \epsilon}{\partial {\bf k}}$
      for this band can be verified to be unchanged by considering
      a Hamiltonian ${\bf h \cdot \tau}$ with ${\bf h}$ linear in ${\bf k}$.]
      This completes the proof, provided
      all bands are accounted for.


\begin{thebibliography}{plain}

\bibitem{Kim13} H-H. Kim et al, Phys. Rev. Lett. {\bf 111} 246603

\bibitem{Zhang15} C. Zhang et al, arXiv:1503.02630
\bibitem{Huang15} X. Huang et al, arXiv:1503.01304


\bibitem{Xiong15} X. Xiong et al, arXiv:1503.08179

\bibitem{Li15} C.-Z. Li et al, arXiv:1504.07398



\bibitem{SS13} D. T. Son and B. Z. Spivak, Phys. Rev. B {\bf 88}, 104412 (2013)

\bibitem{NM83} H. B. Nielsen and M. Ninomiya, Phys. Lett.  {\bf 130B}, 389 (1983)

\bibitem{Kim13b} Y.-S. Jho and K.-S. Kim, Phys. Rev. B {\bf 87}, 205133 (2013)

\bibitem{Kim14} K.-S. Kim, H.-J. Kim and M. Sasaki, Phys. Rev. B {\bf 89}, 195137 (2014)



\bibitem{Burkov14} A. A. Burkov, Phys. Rev. Lett. {\bf 112}, 247203 (2014)

\bibitem{Sundaram99} G. Sundaram and Q. Niu, Phys. Rev. B {\bf 59}, 14915 (1999)

\bibitem{Panati03} G. Panati, H. Spohn and S. Teufel, Commun. Math. Phys. {\bf 242}, 547 (2003)

\bibitem{Xiao05} D. Xiao, J. Shi and Q. Niu, Phys. Rev. Lett. {\bf 95}, 137204 (2005)

\bibitem{Duval06} C. Duval, Z. Horv\'{a}th, P. A. Horv\'{a}thy,
L. Martina and P. C. Stichel, Phys. Rev. Lett. {\bf 96}, 099701 (2006)

\bibitem{Xiao06} D. Xiao, J. Shi and Q. Niu, Phys. Rev. Lett. {\bf 96}, 099702 (2005)

\bibitem{Zhou13} J.-H. Zhou, H. Jiang, Q. Niu and J.-R. Shi,
Chin. Phys. Lett. {\bf 30}, 027101 (2013)

\bibitem{VF13} M. M. Vazifeh and M. Franz, Phys. Rev. Lett. {\bf 111}, 027201 (2013)

\bibitem{AH}
  T. Jungwirth, Q.Niu and A. H. MacDonald, Phys. Rev. Lett. {\bf 88}, 207208 (2002);
Z. Fang {\it et al}, Science {\bf 302}, 92 (2003)

\bibitem{FS} F. D. M. Haldane, Phys. Rev. Lett. {\bf 93}, 206002 (2004);
see also D. Vanderbilt, I. Souza and F. D. M. Haldane,
arXiv:1312.4200

\bibitem{PN} See, e.g., D. Pines and P. Nozi\`{e}res, "The Theory of
Quantum Liquids,  Volume 1: Normal Fermi Liquids", W. A. Benjamin,
New York 1966.

\bibitem{footnoteKim14}  In \cite{Kim14}, the interpocket collision term for
a given ${\bf k}$
is taken to be proportional to $n_{\chi} ({\bf k}) - n_{-\chi}({\bf k})$
at the same ${\bf  k}$,
where $\pm \chi$ denotes the two Weyl branches (see their (29)).   It seems that
this assumption is too drastic physically.
In the limit of long intervalley scattering time, the magneto-conductivity
obtained in \cite{Kim14} (their (58)) has a form different from our (\ref{JSS}).






\end{thebibliography}
\end{document}